\documentclass[preprint
 ,secnumarabic%
,amssymb,amsmath,nobibnotes,aps,pre,superscriptaddress,showpacs,showkeys,floatfix]{revtex4}
\usepackage{bm}%
\usepackage{graphicx}
\expandafter\ifx\csname package@font\endcsname\relax\else
 \expandafter\expandafter
 \expandafter\usepackage
 \expandafter\expandafter
 \expandafter{\csname package@font\endcsname}%
\fi

\begin{document}

\title{Dynamics and thermodynamics of rotators interacting with both long
and short range couplings}

\author{Alessandro Campa}
\email{campa@iss.infn.it}
\affiliation{Complex Systems and Theoretical Physics Unit, Health and Technology Department,
Istituto Superiore di Sanit\`a, and INFN Roma1 - Gruppo Collegato Sanit\`a,
Viale Regina Elena 299, 00161 Roma, Italy}
\author{Andrea Giansanti}
\email{andrea.giansanti@roma1.infn.it}
\affiliation{Physics Department, Universit\`a di Roma ``La Sapienza'', P.le Aldo Moro 2,
00185 Roma, Italy}
\author{David Mukamel}
\email{david.mukamel@weizmann.ac.il}
\affiliation{Department of Physics of Complex Systems, The Weizmann Institute
of Science, Rehovot 76100, Israel}
\author{Stefano Ruffo}
\email{stefano.ruffo@unifi.it}
\affiliation{Dipartimento di Energetica ``Sergio Stecco'' and CSDC,
Universit\`a di Firenze, and INFN,
Via s. Marta 3, 50139 Firenze, Italy}

\date{\today}%

\begin{abstract}
The effect of nearest-neighbor coupling on the thermodynamic and
dynamical properties of the ferromagnetic Hamiltonian Mean Field
model (HMF) is studied. For a range of antiferromagnetic
nearest-neighbor coupling, a canonical first order transition is
observed, and the canonical and microcanonical ensembles are
non-equivalent. In studying the relaxation time of non-equilibrium
states it is found that as in the HMF model, a class of
non-magnetic states is quasi-stationary, with an algebraic
divergence of their lifetime with the number of degrees of freedom
$N$. The lifetime of metastable states is found to increase
exponentially with $N$ as expected.

\end{abstract}


\maketitle

\section{Introduction}\label{intro}

Hamiltonian systems with long-range interactions exhibit several
peculiarities, both in their equilibrium properties and in the
dynamical behavior~\cite{draw}. At equilibrium, it is often found
that different statistical ensembles can be non-equivalent even in
the thermodynamic limit. First discussed theoretically by Hertel
and Thirring~\cite{heth}, this result has been confirmed in
analytical and numerical studies (see, e.g.,
Refs.~\cite{toan,isco,elht,bamr,leru,chom,eltt,coet,bbdr}), which
pointed out that ensemble inequivalence is usually associated with
the presence of a first order phase transition in the canonical
ensemble.

Attention has also been focussed on the dynamics, in particular on
the relaxation to equilibrium of a system initially prepared in a
state which is far from equilibrium. Many studies have been
devoted to the Hamiltonian Mean Field (HMF) model~\cite{anru}, a
system of classical XY-rotators with infinite range ferromagnetic
couplings. This model exhibits a second order transition to a
magnetized state. It has been shown numerically in microcanonical
simulations that for a large class of non-equilibrium initial
conditions the model exhibits fast relaxation to another
non-equilibrium state, in a fashion similar to the violent relaxation
of astrophysical systems~\cite{lybe,padm}. After this transient
the relaxation to Boltzmann-Gibbs equilibrium takes place in times
that diverge with the number $N$ of rotators, according to a power
law~\cite{anru,lrr1,lrr2,pllr}. The stability of these long-lived
states, that have been called ``quasi-stationary'' states, has
recently been analyzed in terms of Vlasov and Fokker-Planck
equations~\cite{yama,chch,ante,chavanis}. It should be emphasized
that these quasi-stationary states are not metastable states in
the thermodynamical sense, since they do not correspond to local
maxima of the entropy (in the microcanonical ensemble) or local
minima of the free energy (in the canonical ensemble). A similar
effect has recently been observed in the Ising model with long and
short range interactions, where the lifetime of the
quasi-stationary states diverges logarithmically with
$N$~\cite{mrsc}.

In this work we consider the robustness of the quasi-stationary
states with respect to perturbations of the long-range
Hamiltonian. This problem is of interest if one wants to ascertain
the relevance of these states in realistic long-range systems,
where a purely long-range Hamiltonian could be perturbed by
external forces or by the presence of additional short-range terms
in the potential energy. In particular, we generalize the HMF
model to include nearest-neighbor interaction between the
rotators, in addition to the long range interaction.

The paper is organized as follows: In Section \ref{model} we
introduce our model; in Section \ref{equilprop} we summarize the
equilibrium properties of the model as obtained within the
canonical and the microcanonical ensembles. In Sections
\ref{quasistat} and \ref{meta} we report our numerical results on the
relaxation time of quasi-stationary and metastable states,
respectively. Concluding remarks are given in  Section
\ref{discuss}.

\section{The model}\label{model}

We study a system of $N$ rotators characterized by phase variables
$\theta_i$, whose Hamiltonian is given by:
\begin{equation}\label{hamil}
H=\sum_{i=1}^N \frac{p_i^2}{2} + \frac{J}{2N}\sum_{i,j=1}^N
\left[1-\cos \left( \theta_i -\theta_j \right) \right] -K
\sum_{i=1}^N \cos \left( \theta_{i+1}-\theta_i \right)~.
\end{equation}
The $N$ rotators are placed on a one-dimensional lattice with
periodic boundary conditions ($\theta_{N+1}\equiv \theta_1$). The
choice of a one-dimensional lattice is made for mathematical
convenience, as it will be explained in the following Section. The
parameter $K$ is the short-range (actually nearest-neighbor)
coupling between rotators. For $K=0$ this system reduces to the
HMF model, that, for positive $J$, has a second order
ferromagnetic transition at the critical temperature $T_c=0.5 J$,
corresponding to the critical energy density $\epsilon_c=0.75
J$~\cite{anru}. Clearly in this case the topology of the lattice
is irrelevant. The transition is characterized by the value of the
order parameter given by the magnetization $m$:
\begin{equation}\label{magdef}
m=\frac{1}{N}\sqrt{\left(\sum_{i=1}^N \cos \theta_i \right)^2 +
\left(\sum_{i=1}^N \sin \theta_i \right)^2} = \sqrt{m_x^2 + m_y^2}.
\end{equation}
In the $N\rightarrow \infty$ limit, $m$ is positive below the
critical energy, approaching zero for $\epsilon \rightarrow
\epsilon_c$; while $m$ is identically zero above $\epsilon_c$. It
has been shown that for the HMF model microcanonical and canonical
ensembles are equivalent ~\cite{bbdr}; this equivalence had been
discussed before for a general class of magnetic systems with long
range interactions \cite{campajpa}. Microcanonical simulations in
which the system is initially prepared in an homogeneous state,
i.e. with the angles $\theta_i$ uniformly distributed between $0$
and $2\pi$ (and thus $m \approx 0$ at time $t=0$) and the $p_i$'s
uniformly distributed between $-p_0$ and $p_0$ (with $p_0$
determined by the energy) show the following: if the energy per
particle is in a certain range below the critical energy, then the
system remains for a long time in this non-magnetic state, and it
approaches the equilibrium magnetized state only after a time that
diverges like $N^{1.7}$~\cite{yama}. In the same subcritical
energy range it is found that if the $\theta_i$'s are initially
all equal, with the $p_i$'s uniformly distributed, there is a fast
relaxation to a non-magnetic state, and again the equilibrium
magnetized state is reached after a time diverging with $N$,
although with a smaller power law~\cite{pllr}.

In the following sections we study the equilibrium properties of
system (\ref{hamil}) for general values of $K$, and consider the
dynamical properties at small values of $K$, when the system can
be thought of as a perturbed HMF.

\section{Equilibrium properties}\label{equilprop}
The canonical partition function and the microcanonical entropy of
the model can be obtained by a straightforward generalization of the
corresponding calculations of the HMF model ~\cite{bbdr,anru}. We
will not report here the full details of these calculations, which
we defer to a longer paper, but rather restrict ourselves to a
brief sketch of the derivation. Without loss of generality we take
$J=1$ in~(\ref{hamil}).

\begin{figure}[!tbp]
\centering
\includegraphics{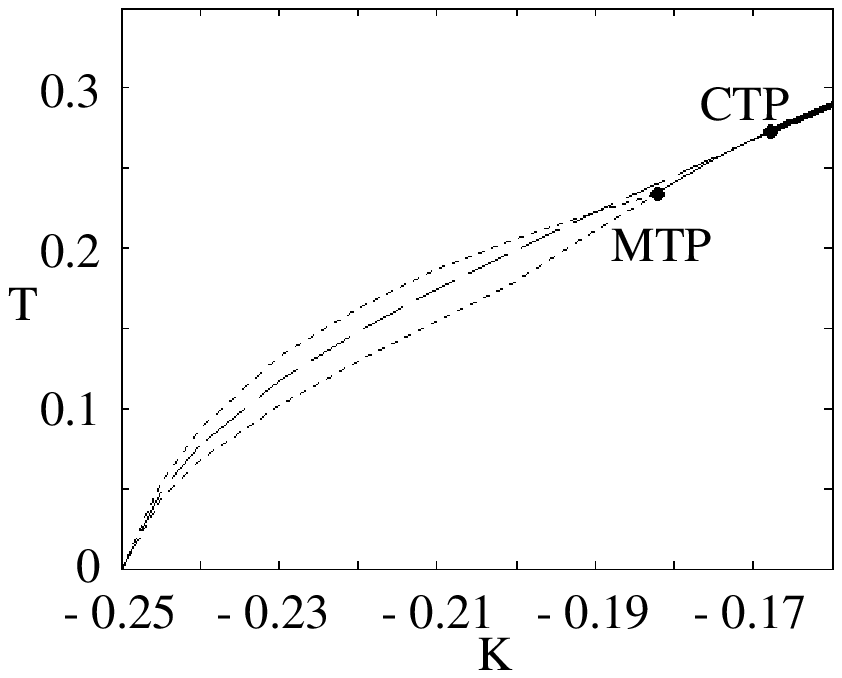}
\caption{
The canonical and microcanonical $(K,T)$ phase diagram. In the canonical ensemble
the transition is continuous (bold solid line) down to
the tricritical point CTP where it becomes first order (dashed
line). In the microcanonical ensemble the continuous transition
coincides with the canonical one at large $K$ (bold solid line). It
persists at lower K (light solid line) down to the tricritical point
MTP where it turns first order, with a branching of the transition
line (dotted lines).
\label{fig1}}
\end{figure}

We begin by considering the free energy per particle $f(\beta)$,
obtained from the calculation of the canonical partition function.
Applying the Hubbard-Stratonovich transformation, $f(\beta)$ can
be written as
\begin{equation}\label{partition}
-\beta f(\beta)=
\max_m \left[ -\frac{\beta (1+m^2)}{2} + \ln \lambda(\beta m, \beta K)
+\ln \frac{2 \pi}{\beta} \right]~.
\end{equation}
In this expression $\lambda(z,\alpha)$ is the largest eigenvalue of the
transfer matrix given by the symmetric integral operator ${\cal T}$ defined as
\begin{equation}\label{transfer}
({\cal T} \psi) (\theta) = \int d \theta' \exp \left[ \frac{1}{2} z (\cos \theta
+\cos \theta') + \alpha \cos (\theta - \theta') \right] \psi (\theta')~.
\end{equation}
The energy density $\epsilon(\beta)$ is computed from (\ref{partition}) using
standard expressions. The microcanonical entropy as a function of
$\epsilon$ can be computed from the argument of maximization in (\ref{partition})
using the recipe introduced in Ref.~\cite{leru}, performing first a minimization over
the inverse temperature $\beta$ before maximizing over the order parameter $m$
\begin{equation}\label{entropy}
s(\epsilon) = \max_m \min_{\beta} \left[ \beta \epsilon + \ln \frac{2 \pi}{\beta}
-\frac{\beta (1+m^2)}{2}+\ln \lambda(\beta m, \beta K) \right]
\equiv \max_m s(\epsilon,m)~.
\end{equation}
The value of $\beta$ corresponding to the solution of the
optimization problem (\ref{entropy}) gives the microcanonical
temperature as a function of $\epsilon$. The phase-diagram in both
the canonical and the microcanical ensembles can be directly
derived from the properties of free energy and entropy. For
positive (i.e., ferromagnetic) nearest-neighbor couplings the
system behaves in a way which is analogous to the HMF ($K=0$)
case: there is a second order ferromagnetic transition (at a $K$
dependent critical temperature and energy), ensembles are
equivalent and the temperature-energy relation (so-called caloric
curve) is qualitatively similar to that of the HMF model. It is
found that there exists a range of negative values of the
short-range coupling $K$ for which the system exhibits a first
order phase transition in the canonical ensemble. This is a result
of the competing effect of the mean-field potential and the
antiferromagnetic short-range coupling. As expected, the canonical
and the microcanonical ensembles are non-equivalent in this range
of coupling. The $(K,T)$ phase diagram of the model is given in
Fig.~\ref{fig1}. For clarity the plot is restricted to the
interesting negative $K$ region.
Starting from positive values and decreasing $K$, the canonical
critical line ends at the tricritical point (CTP) $K=K_1\approx
-0.168$, $T=T_1\approx 0.273$, where the canonical phase
transition becomes first order. The dashed line in Fig.~\ref{fig1}
shows the canonical transition temperature for $K<K_1$. This line
ends at $K=-0.25$, $T=0$. For smaller values of $K$ the system is
disordered at any temperature and there is no transition. This can
be easily understood by computing the energies of the staggered
antiferromagnetic $m=0$ state and that of the fully magnetized
$m=1$ state. For $K \leq -1/4$ the former is always favoured, even
at zero temperature. In the microcanonical ensemble the critical
line extends down to the microcanonical tricritical point (MTP)
located at $K=K_2\approx -0.182$, $T=T_2\approx 0.234$, where the
transition becomes first order. Below this point, a temperature
jump between the two microcanonical temperatures given by the
dotted lines appears. Again, for $K<-0.25$ there is no
transition. The two ensembles are therefore non-equivalent for
$-0.25<K<K_1$.

\section{Quasi-stationary states}\label{quasistat}

\begin{figure}[!tbp]
\centering
\includegraphics{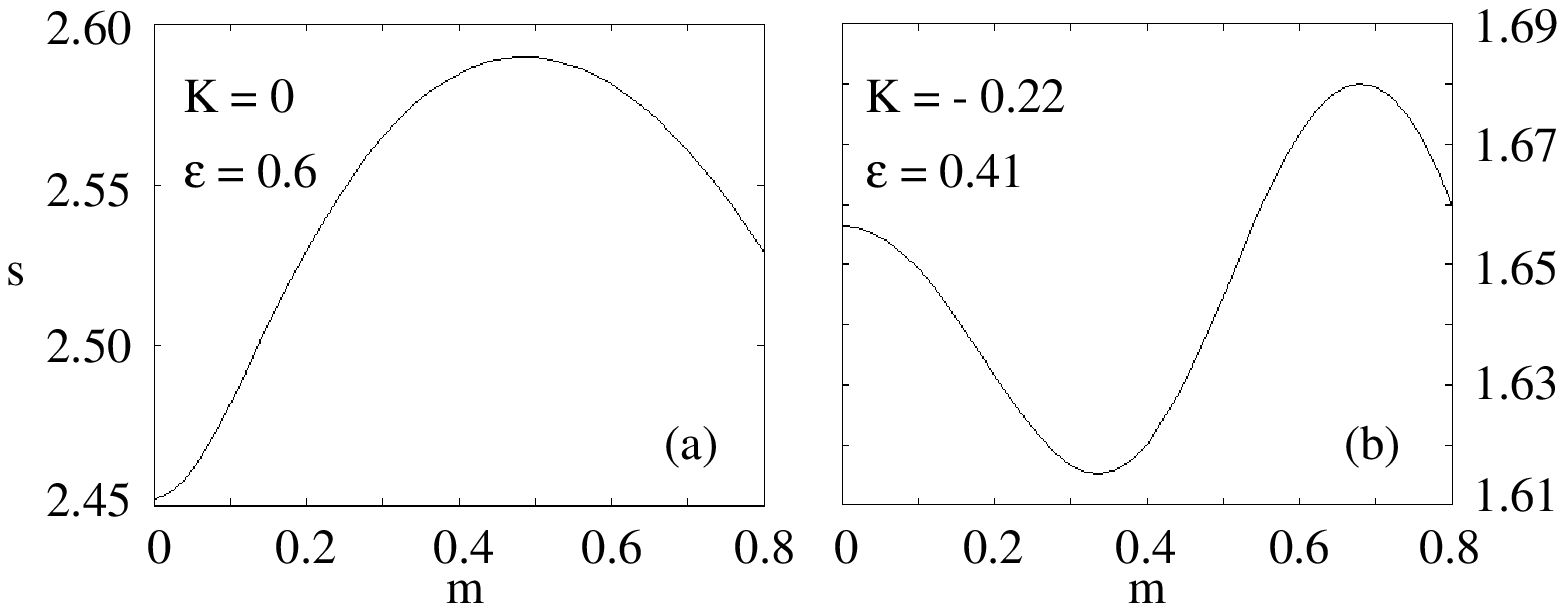}
\caption{Two examples of $s(\epsilon,m)$: (a) Local entropy minimum for a $m=0$
quasi-stationary state; (b) Local entropy maximum for a metastable $m=0$ state.
\label{fig2}}
\end{figure}

The equilibrium magnetization at a given energy $\epsilon$ is
obtained by maximizing the entropy $s(\epsilon,m)$ with respect to
$m$ (see Eq. (\ref{entropy})). Local maxima correspond to
metastable states, that we consider in the next Section. For small
$K$ values (both positive and negative) the entropy dependence on
magnetization shows only a global maximum, at the equilibrium value
of $m$. A typical entropy curve is given by the one shown in
Fig.~\ref{fig2}(a), which has $K=0$ (HMF model); the entropy has a
local minimum at $m=0$.
We are primarily interested in this case since we want to
study the difference in the dynamical behavior induced by a
perturbation of the HMF system, for which the long lifetime of
quasi-stationary states, with $m$ close to $0$, is strictly of
dynamical origin~\cite{yama}. We consider here only the case of
initial homogeneous non-magnetic states, with $\theta$ and $p$
both uniformly distributed.

We performed runs at different $K$ values. For each $K$ we have
chosen a value of the energy density $\epsilon$ below the critical
energy, $\epsilon=\epsilon_c(K)-\Delta \epsilon$, keeping $\Delta
\epsilon=0.06$ fixed. This distance is the same as for the
simulations typically performed at $K=0$ (HMF), where
$\epsilon=0.69$ is studied, with the critical energy density being
at $0.75$~\cite{yama}. We have performed runs for a number of
rotators $N$ ranging from $1024$ to $32768$.

\begin{figure}[!tbp]
\centering
\includegraphics{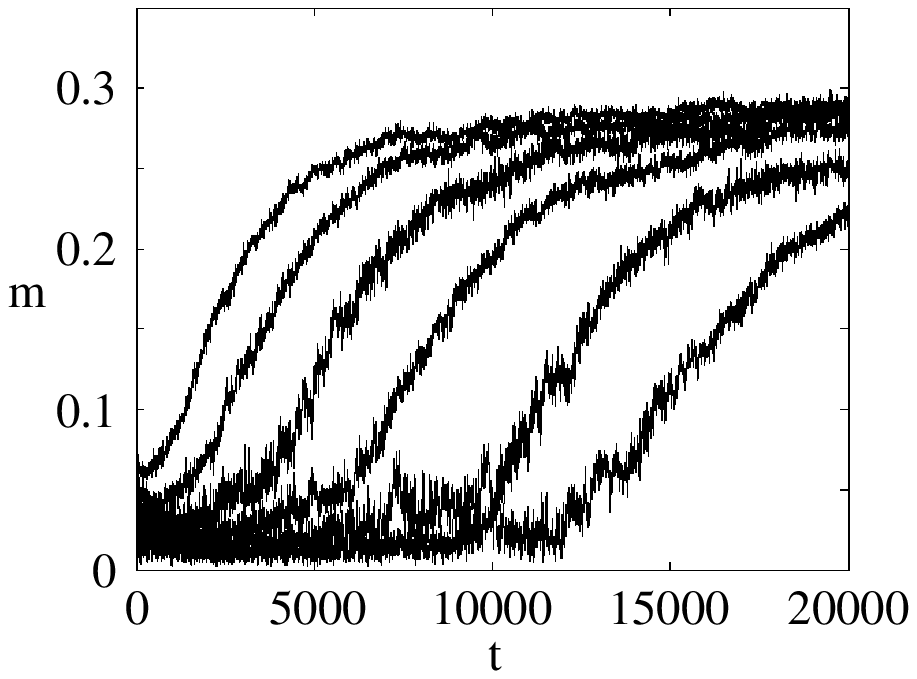}
\caption{Evolution of the magnetization for quasi-stationary states with increasing
number of rotators. Here $K=0.05$ and $\epsilon =0.7115$. From left to right $N=1024$
(100 runs), $N=2048$ (30 runs), $N=4096$ (10 runs), $N=8192$ (10 runs), $N=16384$
(2 runs), $N=32768$ (2 runs).
\label{fig3}}
\end{figure}

\begin{figure}[!tbp]
\centering
\includegraphics{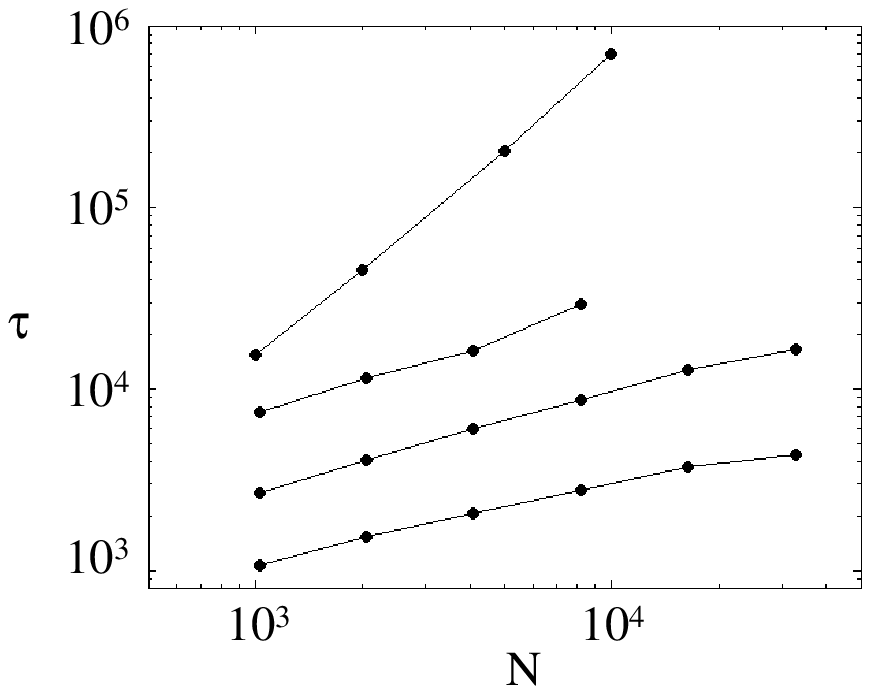}
\caption{Growth of the lifetime of quasi-stationary states with $N$ for different
values of $K$. From top to bottom the data refer to $K=0$, $K=0.0025$, $K=0.05$,
$K=0.1$. Data for $K=0$, with $\tau$ diverging as $N^{1.7}$, are taken from
Ref.~\cite{yama}.
\label{fig4}}
\end{figure}

It turns out that the escape from non-magnetic quasi-stationary
states takes place after a time that fluctuates from run to run
(corresponding to different realizations of the uniform initial
distributions of $\theta$ and $p$), although these fluctuations
are much smaller than those of the case of the thermodynamic
metastable case discussed in the following Section.
In Fig.~\ref{fig3} we plot the instantaneous
magnetization as a function of time for the case $K=0.05$ and
$\epsilon =0.7115$, and for the increasing $N$ values from left to
right. These plots are obtained by averaging over several runs
(the number of these runs for each $N$ value is reported in the
caption). It is worth mentioning that for a given $N$ the lifetime
of the quasi-stationary states is considerably shorter than that of
the HMF model (see, e.g., Ref.~\cite{yama}).
To define this lifetime we have chosen the time at which, during
the rise of the magnetization towards the equilibrium value, the
magnetization has a value equal to the average between the plateau
value in the quasi-stationary state and the equilibrium value. In
order to smooth out fluctuations, we used a running average of the
magnetization over a sliding time window of a conveniently chosen
size. This definition of the lifetime is practically equivalent to
that used in Ref.~\cite{yama} through a more sophisticated fitting
procedure. In Fig.~\ref{fig4} we plot the lifetime $\tau$ as a
function of $N$, in a log-log plot for different $K$ values. It is
evident that the lifetime $\tau$ increases with $N$, possibly
diverging in the limit $N\rightarrow\infty$. This suggests that
quasi-stationarity persists when short range interactions are
added to the HMF model. A naive fit of the curves to the form
$\tau \sim N^\gamma$ with $K$-dependent power-law exponents
$\gamma(K)$ yields $\gamma(0.0025) \approx 0.65$, $\gamma(0.05)
\approx 0.56$, $\gamma(0.1) \approx 0.45$. Thus the observed
exponent $\gamma$ decreases with increasing $K$. For $K>0.1$ we do
not observe a plateau region in the magnetization at $m \approx 0$
and one cannot reasonably speak of quasi-stationarity. Clearly one
cannot rule out the possibility that the seemingly $K$ dependence
of $\gamma$ is just a finite size effect, which are evidently
present in the up-bending of the smaller $K$ value curve in
Fig.~\ref{fig4} and the down-bending of the other two. More
extensive numerical simulations are necessary in order to assess
the law of divergence with $N$.

\section{Metastable states}\label{meta}
\begin{figure}[!tbp]
\centering
\includegraphics{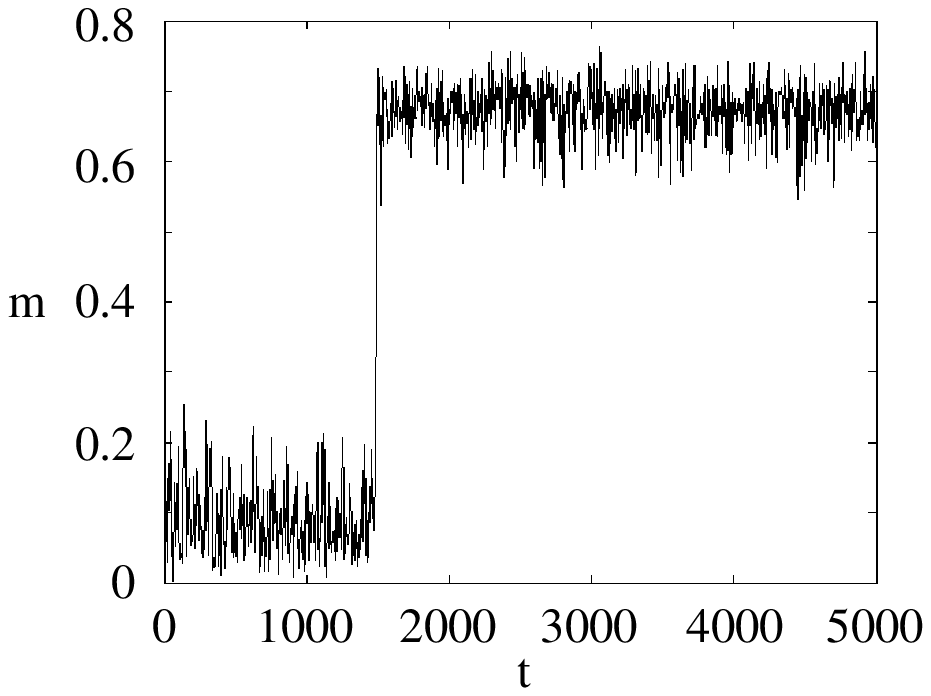}
\caption{Evolution of the magnetization for the metastable situation in which the system with $\epsilon$
and $K$ as in Fig. 2(b) is started with $m=0$.
\label{fig5}}
\end{figure}

\begin{figure}[!tbp]
\centering
\includegraphics{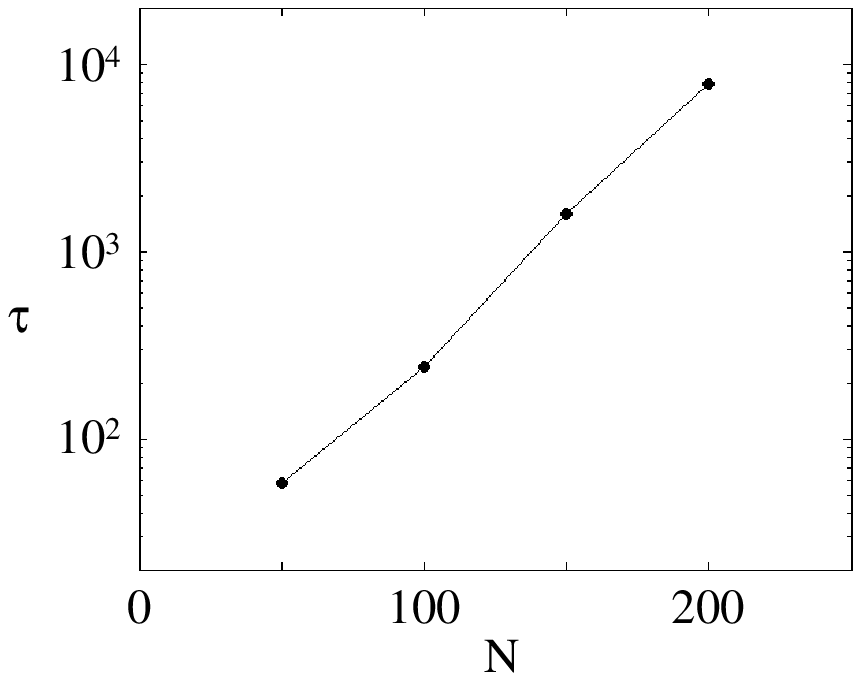}
\caption{Exponential growth of the lifetime of the metastable state of Fig. 3 with the
number of rotators.
\label{fig6}}
\end{figure}

In our system, thermodynamic metastable states
are found for negative $K$ values, for which the
microcanonical transition is first order. In Fig.~\ref{fig2}(b) we
show an example of an entropy function with a local maximum at
$m=0$, while the global maximum is at $m \neq 0$. The evolution of
this $m=0$ state was studied.
In Fig.~\ref{fig5} we show a typical plot of the istantaneous
magnetization $m$ as a function of time, for a system with $N=150$
rotators. After a transient time spent in the metastable state,
there is a sudden transition to the equilibrium magnetization.
In Fig.~\ref{fig6} we plot the lifetime $\tau$ of the metastable state
as a function of $N$. It should be pointed out that an average over many runs is
necessary in order to obtain a sufficiently accurate determination of this value,
since the differences from run to run are not small.
We find that $\tau \sim e^{N \Delta s}$, where $\Delta s$ is the ``entropy barrier'', i.e.
the difference between the entropy at $m=0$ and the one at the minimum (see Fig.~\ref{fig2}(b)).
Thus the transition is determined by the fluctuations at finite $N$, and the exponential
divergence of the lifetime is governed by the entropy barrier. Analogous results
have been found for other systems~\cite{anrt}.

\section{Discussion}\label{discuss}

The model studied in this work offers another example of a
long-range system with inequivalence of ensembles. When the
long-range and the short-range interactions have a competing
character (i.e., when $K$ is negative), we found that the
microcanonical and the canonical ensembles yield different
results, in a range of $K$ values.

The existence of quasi-stationary states seems to be a peculiar
feature of long-range systems, and in this work our aim was to
study their robustness with respect to perturbations of the
long-range nature of the interaction. We found that if the
perturbation is sufficiently small the divergence with $N$ of the
lifetime of these states is preserved. This fact was not obvious a
priori, since one could have argued that even small perturbations
could have given rise to finite lifetimes also in the
$N\rightarrow \infty$ limit. One could have hypothesized the
finiteness of the lifetime on the basis of the fact that, when the
rotators have also a short-range coupling, the $1/N$ expansion of
the Vlasov equation, suitable for the HMF case~\cite{boda}, is no
longer possible. It has to be inferred that, in spite of this fact,
an approximate description of the system with a kinetic equation
possessing out of equilibrium stable stationary states would still
be possible. Its determination is one of the possible directions
of development of this work. It is worth noting that in the case
of the Ising model with both long and short range interactions,
quasistationarity was found to exist, but with a universal
logarithmic divergence of the lifetime with $N$~\cite{mrsc}.

We have considered here only one class of initial conditions. As
was found in the HMF model, other initial states may behave
differently. It would be interesting to explore the dynamical
behavior of other initial conditions.
The model we have analyzed lives on a 1D lattice, but we believe
that some basic dynamical features observed here should extend to
higher dimensions.

It would of course be interesting to determine the range of
validity, across different models, of the results presented in
this paper; this could be relevant for the study of the dynamical
properties, and in particular for the approach to equilibrium of
open long-range systems, i.e., systems in interaction with external
degrees of freedom. One could suppose that the open
character of the system can be mimicked by a perturbation of the
long-range Hamiltonian. Preliminary investigations of a simple
magnetic system with these characteristics~\cite{baor} give results
with features that are similar to those presented in this work.

\begin{acknowledgments}
D. M. acknowledges financial support by the Israel Science Foundation (ISF);
S. R. acknowledges financial support by the project \textit{Order and
Chaos in Nonlinear Extended Systems} of COFIN03-MIUR.
\end{acknowledgments}

\end{document}